\pdfoutput=1
 \documentclass[
reprint,
prl,
]{revtex4-2}

\usepackage{graphicx}
\graphicspath{{./}}     
\usepackage{dcolumn}
\usepackage{bm}
\usepackage{float}
\newcommand {\VEC} [1] {\boldsymbol{#1}}
\usepackage{dcolumn}
\usepackage{xcolor}

\begin{document}
\preprint{APS/123-QED}

\title{Direct measurement of energy transfer in strongly driven rotating turbulence}

\author{Omri Shaltiel}
\altaffiliation{Racah Institute of Physics, The Hebrew University, Jerusalem 91904, Israel}

\author{Alon Salhov}%
\affiliation {Racah Institute of Physics, The Hebrew University, Jerusalem 91904, Israel}

\author{Omri Gat}%
\affiliation {Racah Institute of Physics, The Hebrew University, Jerusalem 91904, Israel}

\author{Eran Sharon}%
\affiliation {Racah Institute of Physics, The Hebrew University, Jerusalem 91904, Israel}

\date{\today}

\begin{abstract}
A short, abrupt increase in energy injection rate into steady strongly-driven rotating turbulent flow is used as a probe for energy transfer in the system. The injected excessive energy is localized in time and space and its spectra differ from those of the steady turbulent flow. This allows measuring energy transfer rates, in three different domains: In real space, the injected energy propagates within the turbulent field, as a wave packet of inertial waves. In the frequency domain, energy is transferred non-locally to the low, quasi-geostrophic modes. In wavenumber space, energy locally cascades toward small wavenumbers, in a rate that is consistent with two-dimensionsal (2D) turbulence models. Surprisingly however, the inverse cascade of energy is mediated by inertial waves that propagate within the flow with small, but non-vanishing frequency. Our observations differ from measurements and theoretical predictions of weakly driven turbulence. Yet, they show that in strongly-driven rotating turbulence, inertial waves play an important role in energy transfer, even at the vicinity of the 2D manifold. 

\end{abstract}

\maketitle

\paragraph{\label{sec:intro}Introduction}
Understanding the dynamics governing rotating turbulent flows is important in scientific fields such as geophysics, astrophysics and atmospheric sciences \cite{pedlosky1987geophysical, davidson2013turbulence, davidson2015turbulence}. However, certain fundamental processes governing these flows remain unclear \cite{alexakis2018cascades}. In particular, the relation between two competing views, that of quasi-two-dimensional (2D) turbulence and that of inertial-wave turbulence, is not clear \cite{scott2014wave,van2020critical}.

Rotating incompressible fluids are described by the rotating-frame Navier-Stokes equations (RNSE) and the incompressibility condition. In this work, we consider a system rotating at a constant rate $\Omega$ around a vertical axis: $\VEC{\Omega}=\Omega\hat{\VEC{z}}$ ($\hat{{}}$ marks a unit vector). The system's vertical extent is comparable to its extent in the ${x},{y}$ “horizontal” directions, perpendicular to the axis of rotation. Two dimensionless numbers characterize such systems, the Reynolds number, $Re=UL/\nu$ and Rossby number, $Ro=U/(2\Omega L)$, where $U$ and $L$ are the typical  velocity and length scales, and $\nu$ is the kinematic viscosity. Rotating turbulence emerges when $Re \gg 1$, indicating the dominance of nonlinear inertial effects over viscous effects, and $Ro \ll 1$, signifying the dominance of Coriolis acceleration over nonlinear inertial accelerations.

The inviscid linearized RNSE, obtained by dropping the inertial and viscous terms, has solutions of the form of plane inertial-waves \cite{greenspan1968theory}. These waves with frequency $\omega$ and a wave-vector $\VEC{k}$ obey the dispersion relation \cite{disspersion_symmetry}:
\begin{equation}
    \omega = \pm \,2\Omega \cdot \cos (\theta)\ ,\qquad \cos(\theta)=\hat{\VEC{k}}\cdot\hat{\VEC{z}}\ ,
    \label{eq:dispersion_relation}
\end{equation}

i.e., the frequency does not depend on the magnitude of the wave vector, $k=|\VEC{k}|$, but only on the angle, $\theta$, between $\hat{k}$ and the rotation axis ($\hat{z}$). The vertical component of the group velocity is given by:
\begin{equation}
    C_{g,z}(\VEC{k}) = \frac{2\Omega \sin^2{\theta}}{k}
    \label{eq:group velocity}
\end{equation}
Thus, for a given $k$, wave packets centered around horizontal wave-vectors with $\theta \approx \pi/2$ transfer energy vertically with the highest speed.

Inertial waves, have been observed in experiments \cite{bordes2012experimental, cortet2020shortcut, yarom2014experimental,cortetcampagne2015disentangling, duran2013turbulence} and simulations \cite{cambon1997energy,duran2013turbulence,godeferd2015structure} of rotating turbulence decay and build-up. \cite{kolvin2009energy, davidson2006evolution, staplehurst2008structure, bewley2007inertial, morize2005decaying, godeferd2015structure}. Theoretical descriptions of rotating turbulence that take into account interactions have been derived \cite{galtier2003weak, nazarenko2011wave, cambon1997energy, smith1999transfer}; they are typically limited to weak wave turbulence. Indeed, for weakly forced steady turbulence or when the geostrophic flow is suppressed, 3D inertial waves dominate the energy spectrum and energy transfer is mediated by three-wave resonant interactions. Such a behavior was predicted theoretically \cite{galtier2003weak} and first shown in recent experimental work \cite{cortet2020shortcut, cortet2020quantitative}. It is also observed in our system, when driven weakly (see supplemental information \cite{supp_material}, figure S2). Yet, at stronger driving, which is relevant to this work, the mode of energy transfer was changed, leading to dominance of the quasi-geostrophic flow. 

Previous observations indicate that as the rate of rotation increases, the flow field becomes increasingly two-dimensional\cite{alexakis2018cascades}. Under these conditions, some characteristics of the horizontal flow resemble those of 2D (non-rotating) turbulence \cite{yarom2013experimental, smith1999transfer,buzzicotti2018inverse,sen2012anisotropy, baroud2003scaling,lamriben2011direct,campagne2014direct}, including the inverse cascade of energy from small to large scales. These characteristics, along with vertical uniformity of the flow, persist even when energy is injected locally in space \cite{baroud2003scaling}, implying the existence of a mechanism homogenizing energy vertically. Motivated by these observations, models that focus on the energy transfer to the geostrophic component of the flow (flow in the 2D horizontal plane) were derived \cite{di2016quantifying, le2020near, cambon1997energy, bellet2006wave, smith1999transfer, gallet2015exact, nazarenko2011critical}. However, the precise mechanism underlying this homogenization process remains not fully understood.

These diverse results show that rotating turbulence is spectrally heterogeneous, energetically dominated by the geostrophic quasi-2D flow component, that coexists with the inertial-wave-dominated 3D flow \cite{yarom2014experimental,yarom2017experimental, le2020near,le2017inertial,clark2014quantification, cortetcampagne2015disentangling}. Still, the relationship between these components is not well understood. Specifically, the process by which energy is transferred from 3D to quasi-2D modes, and the role that inertial waves play in the strongly-driven turbulence regime, are currently not clear. Due to the flow's three-dimensionality and anisotropy answers to these questions are associated with rates of energy transfer in real space, in the frequency domain and in the wavenumber domain. Measuring transfer rates in steady flows is challenging, and such multidimensional measurements have not been conducted yet.

In this work we perturb a rotating turbulent steady state and measure the evolution of the perturbation within a three-dimensional fluid domain. This measurement allows us to probe the energy transfer rate in the three variables mentioned above. In \cite{salhov2019measurements} we showed that an abrupt and short increase in the energy injection rate into existing turbulence (an injection pulse) generates wave-packets of intense turbulence that propagate for long times. In the current study, we demonstrate for the first time that the spectral components of these wave packets obey the dispersion relation of inertial waves. We observe two distinct processes of energy transfer: The first, which has not been measured before, is a rapid, non-local process. It transfers energy from 3D, high frequency waves to low frequency quasi-2D modes. The second, slower energy transfer, is the inverse energy cascade from short to long-wave modes in the quasi-geostrophic manifold. In spite of its similarity with inverse cascade in 2D turbulence, we observe that this process too, is mediated by inertial waves propagating with a small, but non-zero z component of their wave vector.

\paragraph{Experimental setup}\label{sec:exp}
The experimental setup is detailed in the supplementary material \cite{supp_material}. It is composed of a rotating plexiglass cylinder of 80 cm diameter and 90 cm height, placed on a rotating table ($\VEC{\Omega} =- \Omega_z\hat{ \VEC{z}} $, with a maximum rotation rate of $12.6\,\textrm{rad}/\text{s}$).

The tank is filled with water and covered with a transparent flat lid. Energy is injected at the bottom of the tank by circulating the water through an array of outlets and inlets. The energy injection is concentrated at a central wavelength $2\pi/k_{inj}$, which is a decreasing function of $\Omega$ (see \cite{salhov2019measurements}) down to $\sim5\,$cm at high rotation rates.

Using a vertically scanning horizontal laser sheet, we measure the horizontal (${x}$-${y}$ plain) velocity field, $\mathbf{u}_{\perp}(x,y,z,t)$, inside a $\sim 21\times 21\times 24\text{cm}^3$ volume in the interior of the tank. Spatial resolution is $0.22$cm horizontally and  $0.7$cm vertically, at a rate of $21.4$Hz.
In each experiment, the system is brought to steady state by running it for $\sim 300s$ with an angular speed $9.5 \text{rad}/s < \Omega < 12.5  \text{rad}/\text{s}$ and a constant energy injection rate, obtaining a turbulent flow with $0.006<Ro<0.02$ and $1500<Re<2100$.

We then increase the pumping rate for a duration of 1.5$\,$s, chosen as the time interval ($10.5\,\text{s}<t<12\,\text{s}$), after which the injection strength returns to its previous constant value. We continually measure the velocity field until $t=60\,$s. The total energy of the pulse is less than 10 percent of the tank's total energy. Indeed, watching a video of the energy density field\cite{video_ref2, supp_material}, it is difficult to identify the moment of pulse injection. Each experiment is repeated six times, and the measurements are averaged over the repetitions. 

\begin{figure}[b]
    \centering
    \includegraphics[width=0.5\textwidth]{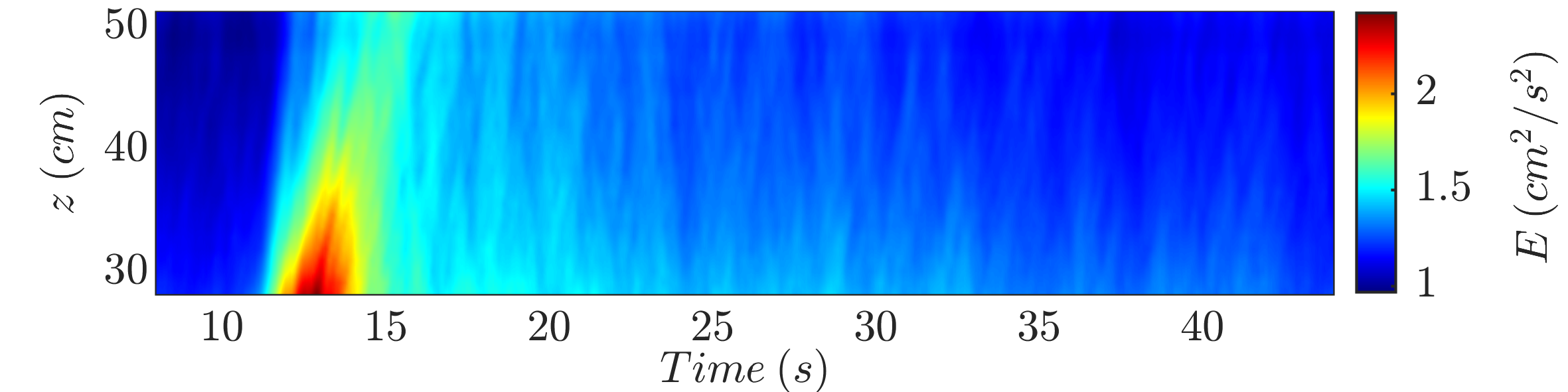}
  
   \caption{
    Horizontally averaged energy density $E(z,t)$ for an ensemble average of 6 experiments with $\Omega=4\pi\,\text{rad}/\text{s}$. Pulse injection occurs at $t=10.5\text{s}$. 
    }
    \label{fig:pulse ilustration}
\end{figure}

The short increase in energy-injection generates a pulse of fluid flow with an enhanced energy density at the bottom of the tank. The pulse propagates upwards on the background of the steady-state turbulent flow (Fig.~\ref{fig:pulse ilustration}). 
During its propagation, the pulse gradually broadens and decays, exciting the entire measurement volume. The flow returns to its steady state after approximately 30 seconds.

We now analyze the spectral properties of the pulse as it evolves in time. 
To confirm that the excess energy in the pulse is composed of inertial waves, we calculated the spectral energy distribution by applying a spatio-temporal Fourier transform to the velocity field. By subtracting the steady state energy density spectrum, we were able to project the excess energy onto the frequency-polar angle plane defined by the dispersion relation Eq.~(\ref{eq:dispersion_relation}). Our results demonstrate that the excess energy is concentrated around the dispersion relation, providing strong evidence that these waves are indeed inertial waves (see supplemental video and supplementary material\cite{video_ref, supp_material}).

\paragraph{\label{sec:level2}Frequency component analysis}
We now calculate the frequency-filtered velocity field $\mathbf{u}_\perp(t,\mathbf{r};\omega)$, applying a Gaussian band-pass filter of width $\sim0.1\Omega$ centered at $\omega$ to the measured velocity field through. 

\begin{figure}
    \centering
    \includegraphics[width=0.48\textwidth]{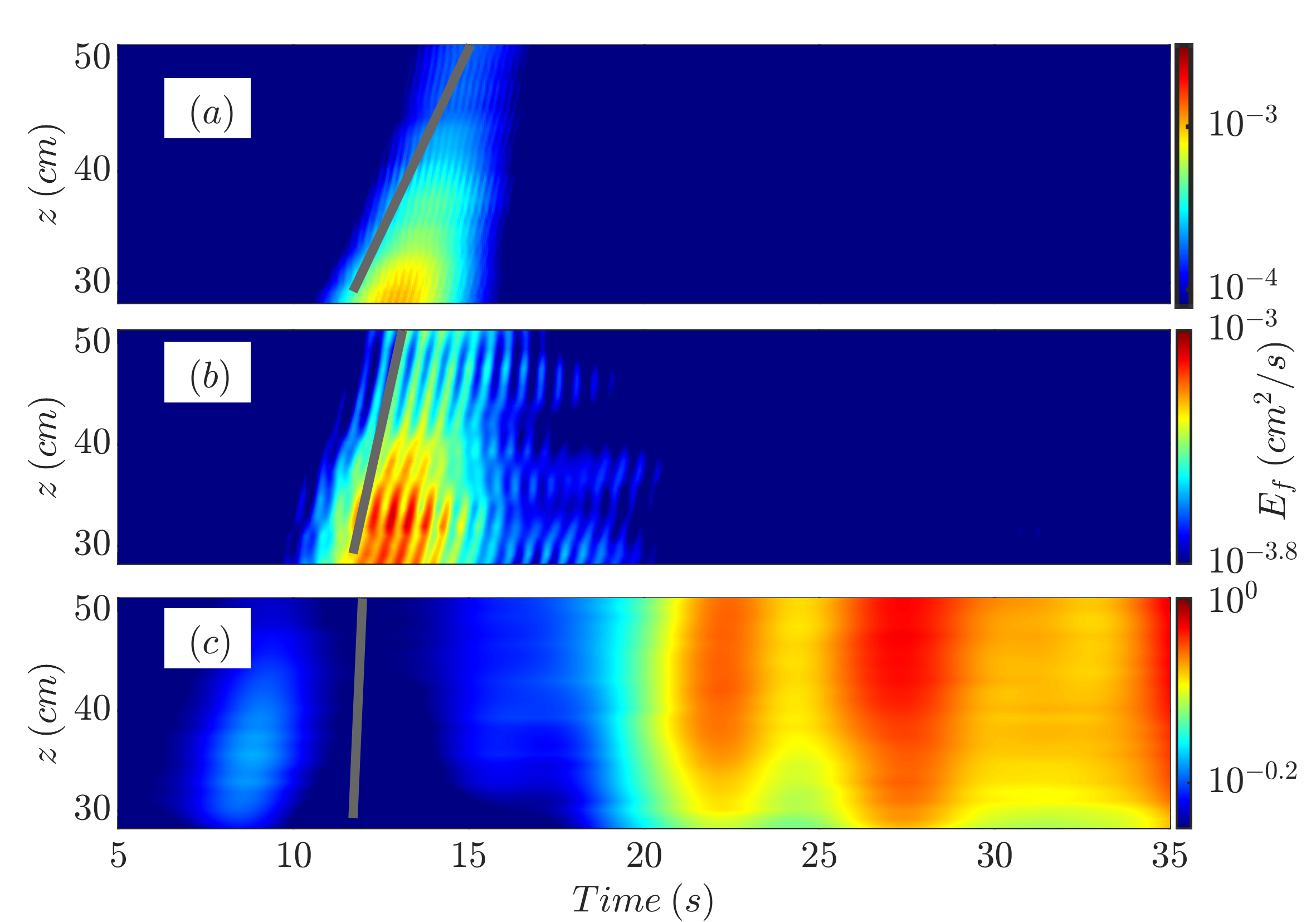}
    \includegraphics[width=0.48\textwidth]{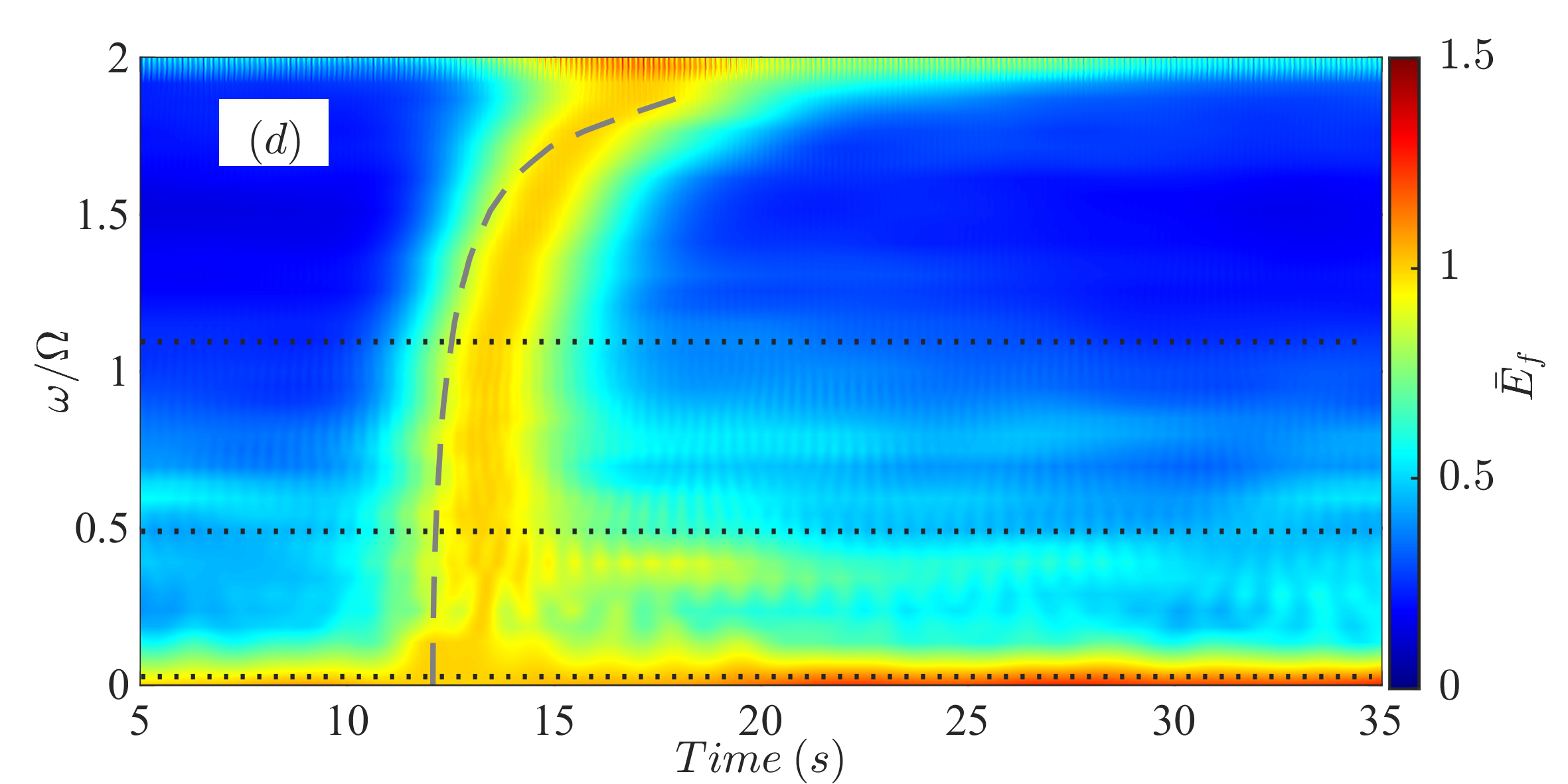}

    \caption{Time evolution of frequency-filtered energy density.
    (a-c) Horizontally averaged energy density $E_f$ over time and vertical position $z$, for $\omega=1.5\Omega(a)$, $0.6\Omega(b)$, and $0.1\Omega(c)$. Gray lines represent to the vertical group velocity of an inertial wave packet with the respective central frequency. The data was collected for experiments with $\Omega=3.5\pi$rad/sec $Re\approx2100$ and $Ro\approx0.006$.
    (d) Spectrogram of fully spatially averaged energy density $\bar E_f$ over time and frequency $\omega$.  The energy of each frequency is normalized to equal 1 at the peak of the pulse. Gray dashed curve marks the theoretical arrival time of an inertial wave packet with frequency $\omega$ to the lowest measured plane.
    }
    \label{fig:omega dynamics}
\end{figure}

The kinetic energy density of the filtered field is $|\mathbf{u}_\perp(t,\mathbf{r};\omega)|^2$ and we average it laterally in order to obtain $E_f(t,z;\omega)$ - the energy contained in a given frequency at a given height and time. It is plotted as a function of $z$ and $t$ for three different values of $\omega$ in Fig.\ \ref{fig:omega dynamics} (a-c). The high-frequency components (panels a,b) propagate vertically with a finite speed that agrees with the vertical group velocity Eq.\ (\ref{eq:group velocity}) for the given frequency, $\omega$ and the injection wave-number $\sim 1.28\,\text{rad}/\text{cm}$, marked by the slopes of the gray lines in Fig.\ref{fig:omega dynamics}. These modes lose their energy within a few seconds. Contrarily, the slow, quasi-geostrophic mode (panel c) does not propagate as a sharp front and its energy increases concurrently with the decay of the high-frequency modes. This suggests a nonlinear process of energy transfer in the frequency domain.

In order to gain insight to this process, we plot a spectrogram of the energy in the time domain,$\bar{E}_f(t;\omega)$ (Fig.\ \ref{fig:omega dynamics} (d)). It is obtained via averaging $E_f(t,z;\omega)$ along z (full spatial average) together with normalization \cite{supp_material}.

The wave packet propagation is revealed by the arrival time of the energy pulse at the lowest measured plane $28\,$cm above the injection plane. This dependence on $\omega$ is consistent with the predicted arrival time marked by the dashed gray line (derived from Eq.\ \ref{eq:group velocity}). This is an additional confirmation for the existence of the linear process of wave propagation, as observed above for the two frequencies studied in Fig.\ \ref{fig:omega dynamics} (a,b), marked by the upper horizontal dotted lines on the spectrogram.
Another key observation is that energy is transferred from the high-frequency \emph{directly} to the low-frequency components. The rapid decay of the energy pulse, which is clearly evident for all frequency components higher than $0.1\Omega$ occurs concurrently with the increase in the energy of the quasi-geostrophic, low-frequency modes. There is no indication for a cascade, or a gradual transfer in the frequency (orientation) domain.
The rapid decay of the high frequency modes occurs within a time-scale of $\tau\approx5$s. Notably, this time-scale is consistent with indirect estimation of the dominant nonlinear time obtained by Yarom \cite{ yarom2017experimental} for strongly-driven steady state flows. There, $\tau$ was deduced from the width of the energy spectrum. Performing similar analysis for our steady flows yields $\tau\sim4$s. 

Since the high-frequency part of the flow is dominated by inertial waves, while the low-frequency flow is quasi-geostrophic, we conclude that the energy injected by the pulse is spatially distributed by linear wave propagation. This is followed by a rapid, direct transfer of energy from modes with significant vertical wave vector components to quasi-geostrophic modes. We conjecture that this rapid process works analogously in the steady flow, forming the mechanism by which energy is fed from inertial waves and accumulates in the quasi-geostrophic flow.

\paragraph{\label{sec:level2}Spatial spectral analysis}

Next, we focus on the evolution of the energy density in wave-number. For this purpose we calculate the horizontal Fourier transform of the energy density $E(z,k_{2d},t)$, and its vertical average $\bar E(k_{2d},t)$. Axial symmetry implies that the mean energy density depends only on the magnitude of the (horizontal) wave vector $\mathbf{k}_{2d}$. The excess energy density due to the pulse $\Delta E$ and $\Delta\bar E$ is obtained from $E$ and $\bar E$ (respectively) by subtracting the mean steady-state energy-density.  

\begin{figure}[t]
    \includegraphics[width=0.49\textwidth]{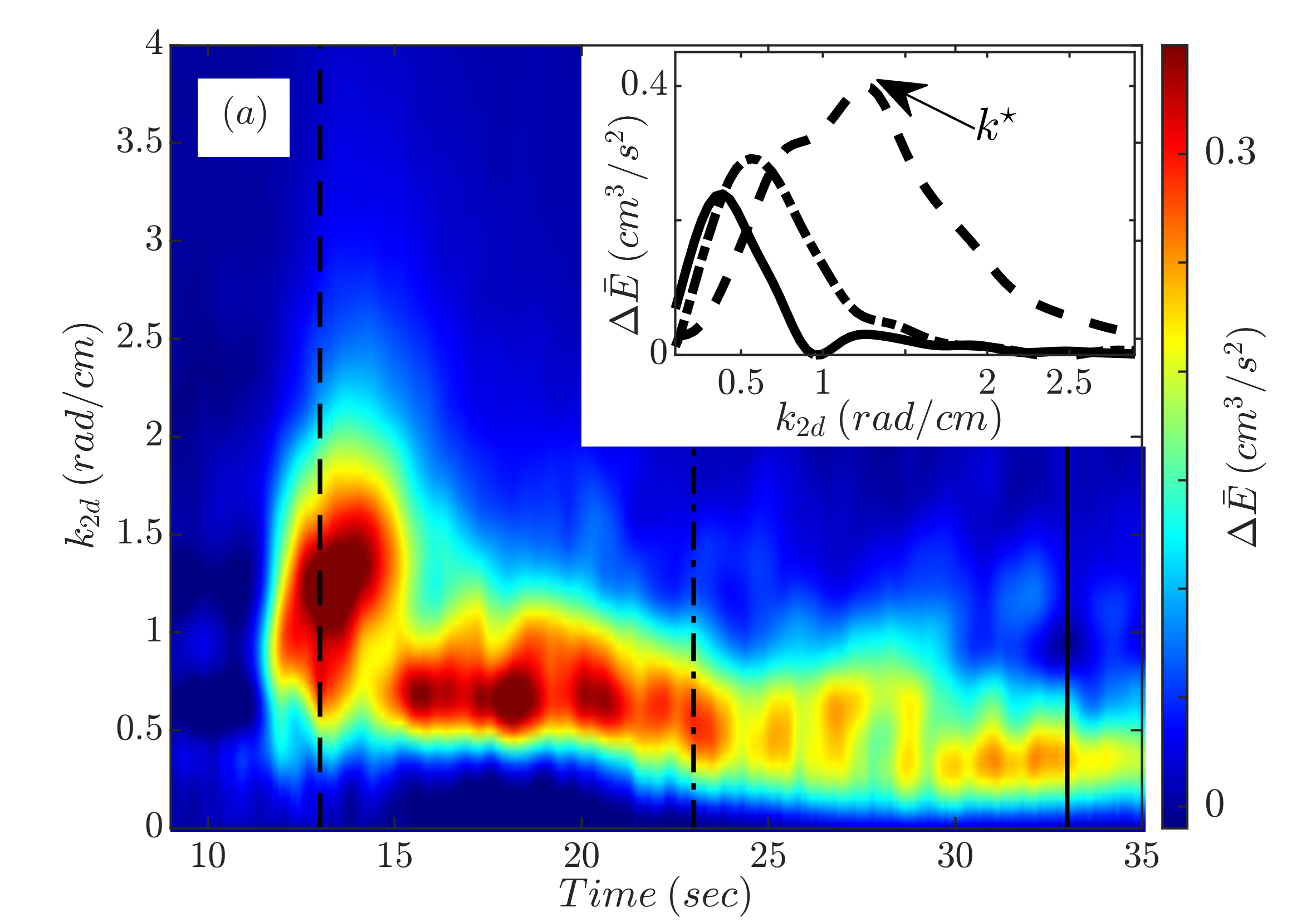}
    \includegraphics[width=0.49\textwidth]{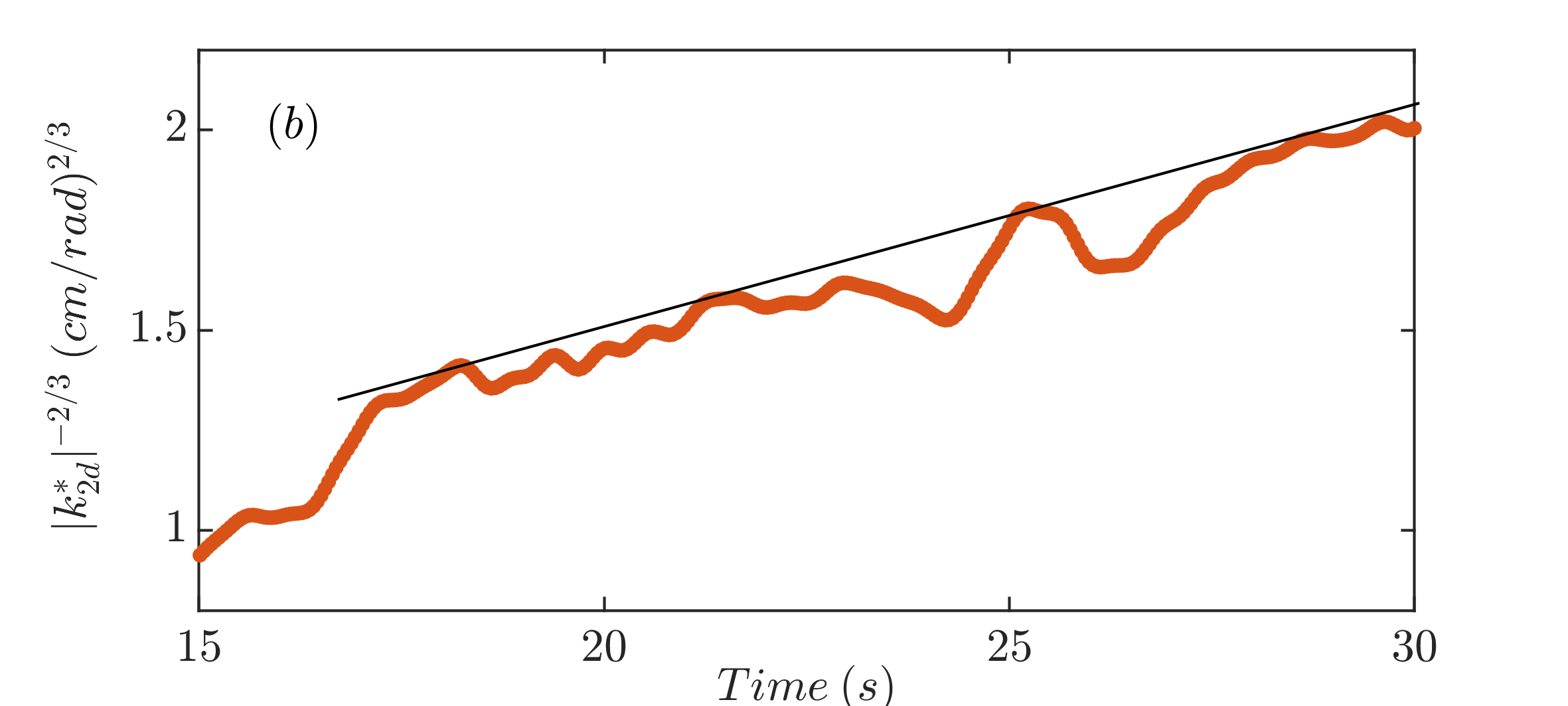}
    \caption{(a) Vertically averaged excess energy density spectrum $\Delta\bar E$, as a function of time and horizontal wave-number $k_{2d}$. Inset: $\Delta\bar E$ as a function of $k_{2d}$ for $t= 13s$ (dashed line), $23s$ (dot-dashed line) and $33s$ (Solid line) (marked on the main figure). The maximum of the instantaneous $\Delta\bar E$, $k^\star_{2d}(t)$, is indicated for $t=13s$. The gradual shift of the excess energy to lower wave-numbers is a manifestation of the inverse energy cascade.
    (b) $|k^\star_{2d}|^{-2/3}$ as a function of time. The agreement with linear growth beyond $t=17s$ indicates $k^\star_{2d}\sim t^{-2/3}$. The data set is the same as in panel (a).
    }
    \label{fig:k figure space avaraged}
\end{figure}

Experimental measurements of $\Delta\bar E$ are presented in Fig.\ \ref{fig:k figure space avaraged} (a) as a function of $t$ and $k_{2d}$, with the inset showing $\Delta\bar E$ as a function of $k_{2d}$ for three different times. Notably, the excess energy distribution in the quasi-geostrophic manifold gradually shifts to smaller wave-numbers (while the total excess energy decreases as the pulse is dissipating). The observations are consistent with a local transfer of energy from short to long-wave modes.

We quantitatively verify this scenario by following the time evolution of $k_{2d}^\star(t)$, the wave-number of the maximum of the averaged excess energy density (Fig.\ \ref{fig:k figure space avaraged} (b)). The linear growth of $|k_{2d}^\star (t)|^{-2/3}$ is consistent with the filling by a constant energy flux of a $k_{2d}^{-5/3}$ spectrum, that characterizes 2D turbulence \cite{kraichnan1967inertial, frisch1984numerical, tabeling2002two,paret1997experimental}, and was also observed in the energy spectrum of the geostrophic component of rotating turbulence \cite{yarom2013experimental}.

\begin{figure}
    \centering
    \includegraphics[width=0.5\textwidth]{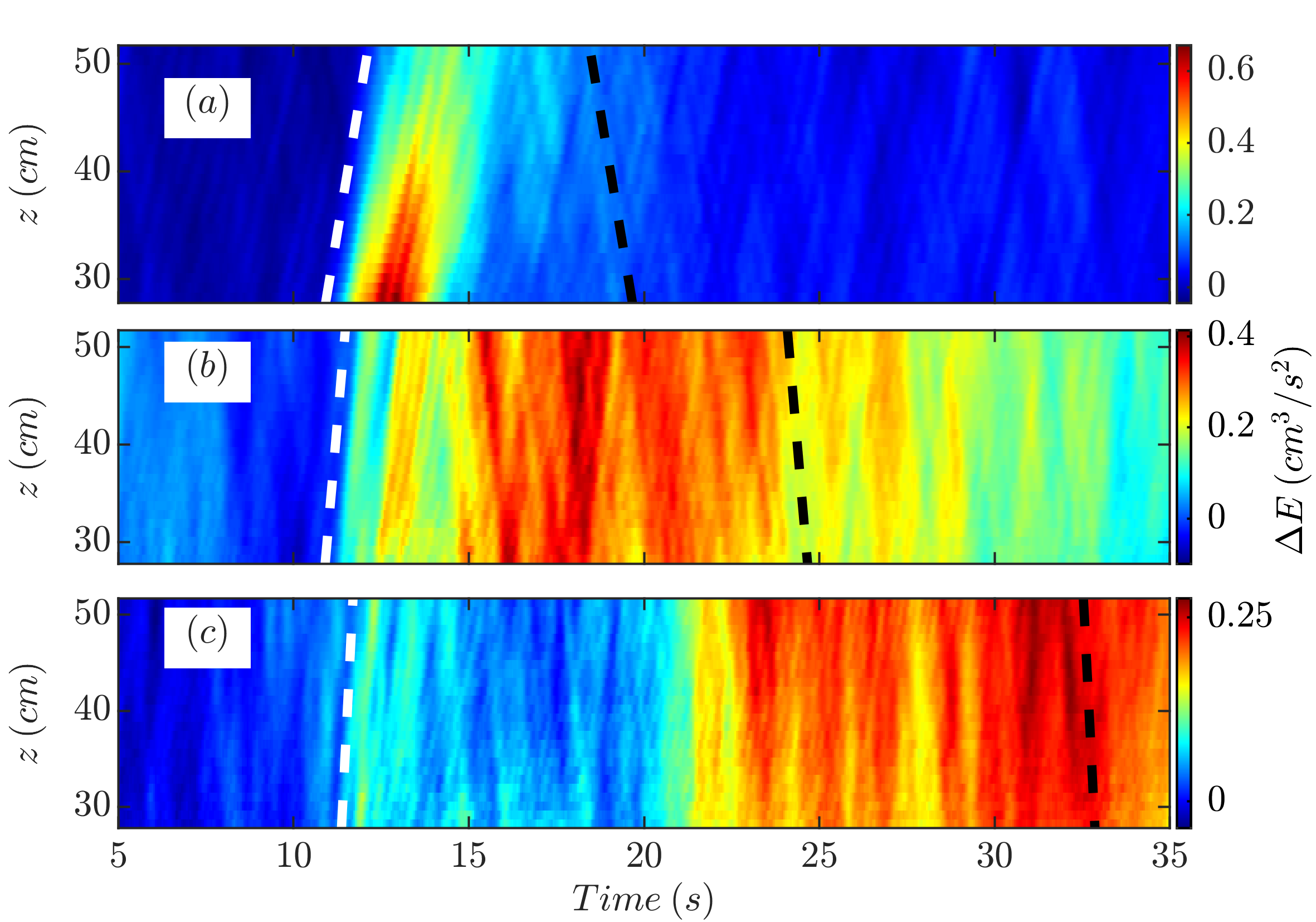}
    \includegraphics[width=0.5\textwidth]{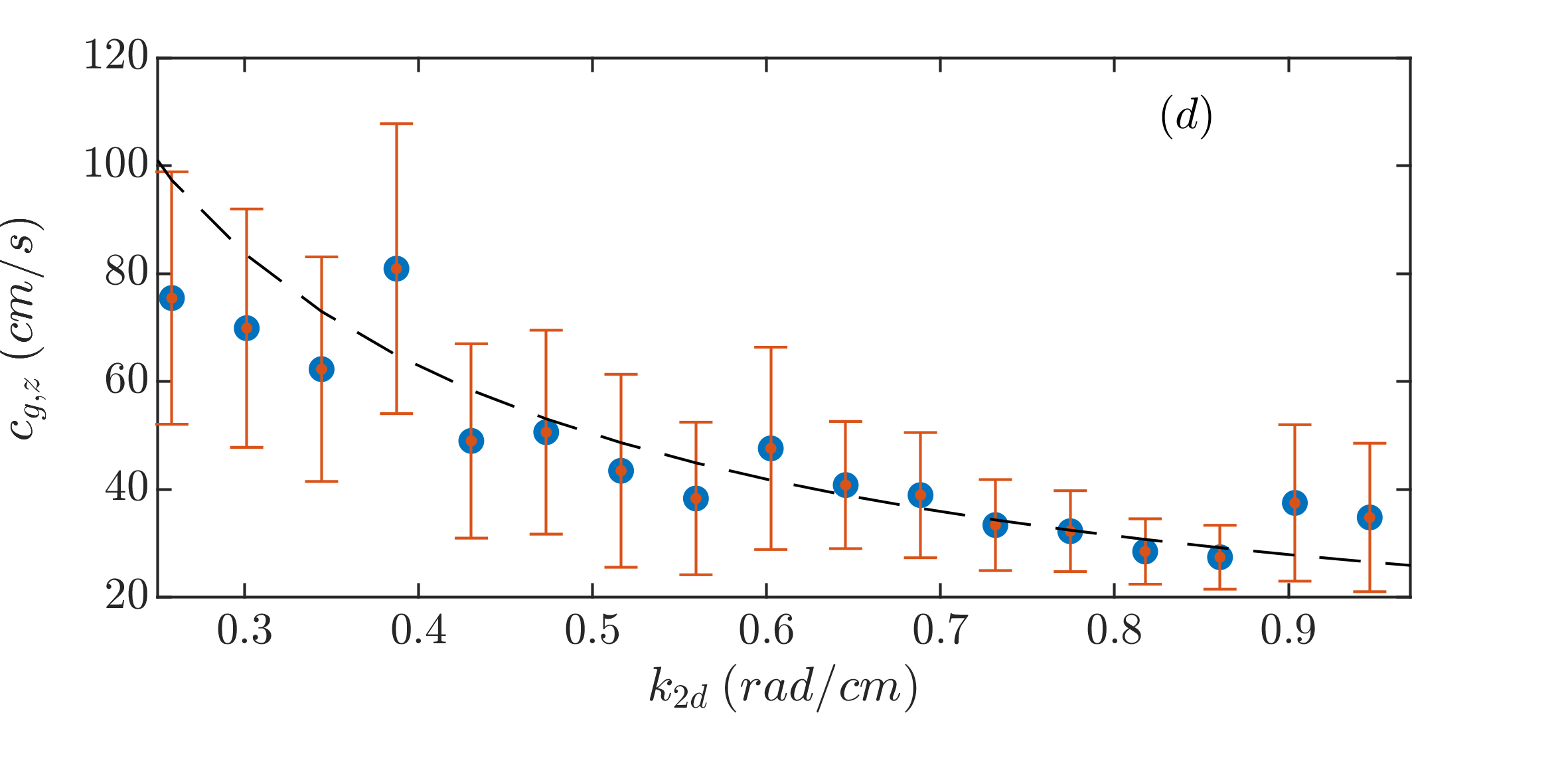}

    \caption{
    (a-c) Excess energy density at fixed wave-number, $\Delta E(z,k_{2d},t)$, as a function of vertical position $z$ and time (showcasing the same data as Figure 3, with vertical resolution). The fixed wave-numbers in each subplot are: $k_{2d}=2.28\text{(a),}\ 0.6\text{(b), and }0.34\text{(c), }  \text{rad}/\text{cm}$. Slopes of the dashed lines indicate $C_{g,z}$ for the relevant wavenumber and $\theta=\pi/2$ for waves that propagate upwards (white) and downwards (black). 
    As time increases, energy cascades from large to small wave-numbers while propagating vertically up and down.
    (d) Estimation of the energy propagation speed for each wave-number $k_{2d}$ (symbols), compared with $C_{g,z}(k_{2d})$ from Eq.\ (\ref{eq:group velocity}) (dashed curve).}
    \label{fig:group velocity as a function of k}
\end{figure}

These observations seem to indicate that the physics of the geostrophic component of rotating turbulence is analogous to that of 2D turbulence. However, our measurement of the $z$-dependent energy density, reveals a weak but significant vertical variation in $\Delta E(z,k_{2d},t)$. Indeed, the plots of $\Delta E(z,k_{2d},t)$ as a function of $z,\,t$ for three fixed values of $k_{2d}$ (Fig.\ \ref{fig:group velocity as a function of k} (a--c)) exhibit \textit{diagonal} rather than vertical correlations, indicating that information travels vertically at a finite speed, while transferring energy to broader horizontal scales. Moreover, the propagation speed is given by the vertical group velocity component (\ref{eq:group velocity}), evaluated at $k_{2d}$, with $\theta=\pi/2$. These waves propagate upwards (white lines) as well as downwards (black lines). Evidence for inertial wave carrying the \emph{excess} energy persists during the entire inverse energy cascade from the large wave-numbers in Fig.\ \ref{fig:group velocity as a function of k} (a), through intermediate (b) to small wave-numbers (c). These observations provide strong evidence that the energy that was shown to inversely cascade in Fig \ref{fig:k figure space avaraged} consists of (nearly horizontal) inertial waves.

Using a Hough transform of the measured $\Delta E(z,k_{2d},t)$ for each $k_{2d}$ \cite{supp_material} we show that the vertical propagation velocity is consistent with the $k$ dependence of $C_{g,z}$ (Fig.\ \ref{fig:group velocity as a function of k} (d)).

\paragraph{\label{sec:level1}Discussion}
We used an injection pulse---a perturbation, localized in time and space, of the energy injection, in order to monitor energy transfer rates in developed rotating turbulence. Assuming that the kinetics that govern the pulse are similar to those of the statistically stationary steady state flow, these measurements facilitate experimental observation of three energy transfer processes: spatial homogenization, transfer of energy from high frequency 3D modes to low-frequency quasi-geostrophic modes, and the inverse cascade of energy within the quasi-geostrophic manifold, which carries the majority of energy in steady state.

The spatial homogenization of energy is achieved by inertial wave propagation, which is the fastest process in the system, and is well-described by linear inertial wave theory, even though the waves are strong and propagate on the background of a turbulent flow.

The injected energy is dominated by inertial waves with 3D wave-vectors. However, the resulting 3D flow field is a short-lived transient, whose energy is siphoned away towards the geostrophic flow. The depletion of the 3D modes occurs on a dynamical timescale determined by interaction of inertial waves with the geostrophic flow \cite{di2016quantifying,yarom2017experimental,salhov2019measurements}.

Energy flows directly from wave-vectors with arbitrary polar angles to quasi-geostrophic wave-vectors, bypassing intermediate ones (Fig.\ref{fig:omega dynamics}). This strongly indicates that the energy transfer in the frequency (orientation) domain is not mediated by intermediate interactions or a cascade.

Once the pulse energy has reached the quasi-geostrophic manifold, it gradually flows to smaller wave-numbers in a manner consistent with an inverse cascade of energy. This observations seems consistent with dynamics analogous to the Kraichnan cascade of 2D turbulence \cite{kraichnan1967inertial,  tabeling2002two}. However, an analysis of the vertically resolved horizontal Fourier transform of the flow shows conclusively that the inverse cascade in Fourier space takes place in parallel with spatial propagation consistent with the linear dynamics of inertial waves. This shows that inertial waves participate in all known energy transfer processes, making the analogy between the rotating turbulence inverse cascade the Kraichnan cascade tenuous.

Some of our results are directly related to theoretical models of rotating turbulence. In this strongly-driven regime we find two nonlinear processes with different time-scales: rapid flattening and a slow inverse cascade, such observations are qualitatively consistent with theoretical predictions \cite{ nazarenko2011critical, le2020near, smith1999transfer}. However, our data exclude energy cascade in the frequency domain (or orientation). This unambiguous result contradicts assumptions incorporated in some of theoretical models\cite{alexakis2018cascades,campagne2014direct}. In addition, to the best of our knowledge, the mediation of the inverse energy cascade by propagating waves was not discussed in theoretical modeling.

We are, therefore, left with several open questions that call for further work: Do the observed propagating modes play a central role in the inverse energy cascade, or are they 3D "artifacts" of the inherently 2D process? Is the infinite (in $z$) medium case a singular limit of the problem? It is possible that the separation to 2D and 3D flow components, as well as the focusing on the exact 2D manifold, are not well determined in finite systems. We hope that further theoretical and experimental works will shed light on these important issues.

\begin{acknowledgments}
This research was supported by the Israel Science Foundation Grant $\#$ 2437/20.
\end{acknowledgments}

\bibliography{main}

\providecommand{\noopsort}[1]{}\providecommand{\singleletter}[1]{#1}%
\begin{thebibliography}{46}%
\makeatletter
\providecommand \@ifxundefined [1]{%
 \@ifx{#1\undefined}
}%
\providecommand \@ifnum [1]{%
 \ifnum #1\expandafter \@firstoftwo
 \else \expandafter \@secondoftwo
 \fi
}%
\providecommand \@ifx [1]{%
 \ifx #1\expandafter \@firstoftwo
 \else \expandafter \@secondoftwo
 \fi
}%
\providecommand \natexlab [1]{#1}%
\providecommand \enquote  [1]{``#1''}%
\providecommand \bibnamefont  [1]{#1}%
\providecommand \bibfnamefont [1]{#1}%
\providecommand \citenamefont [1]{#1}%
\providecommand \href@noop [0]{\@secondoftwo}%
\providecommand \href [0]{\begingroup \@sanitize@url \@href}%
\providecommand \@href[1]{\@@startlink{#1}\@@href}%
\providecommand \@@href[1]{\endgroup#1\@@endlink}%
\providecommand \@sanitize@url [0]{\catcode `\\12\catcode `\$12\catcode `\&12\catcode `\#12\catcode `\^12\catcode `\_12\catcode `\%12\relax}%
\providecommand \@@startlink[1]{}%
\providecommand \@@endlink[0]{}%
\providecommand \url  [0]{\begingroup\@sanitize@url \@url }%
\providecommand \@url [1]{\endgroup\@href {#1}{\urlprefix }}%
\providecommand \urlprefix  [0]{URL }%
\providecommand \Eprint [0]{\href }%
\providecommand \doibase [0]{https://doi.org/}%
\providecommand \selectlanguage [0]{\@gobble}%
\providecommand \bibinfo  [0]{\@secondoftwo}%
\providecommand \bibfield  [0]{\@secondoftwo}%
\providecommand \translation [1]{[#1]}%
\providecommand \BibitemOpen [0]{}%
\providecommand \bibitemStop [0]{}%
\providecommand \bibitemNoStop [0]{.\EOS\space}%
\providecommand \EOS [0]{\spacefactor3000\relax}%
\providecommand \BibitemShut  [1]{\csname bibitem#1\endcsname}%
\let\auto@bib@innerbib\@empty
\bibitem [{\citenamefont {Pedlosky}\ \emph {et~al.}(1987)\citenamefont {Pedlosky} \emph {et~al.}}]{pedlosky1987geophysical}%
  \BibitemOpen
  \bibfield  {author} {\bibinfo {author} {\bibfnamefont {J.}~\bibnamefont {Pedlosky}} \emph {et~al.},\ }\href@noop {} {\emph {\bibinfo {title} {Geophysical fluid dynamics}}},\ Vol.\ \bibinfo {volume} {710}\ (\bibinfo  {publisher} {Springer},\ \bibinfo {year} {1987})\BibitemShut {NoStop}%
\bibitem [{\citenamefont {Davidson}(2013)}]{davidson2013turbulence}%
  \BibitemOpen
  \bibfield  {author} {\bibinfo {author} {\bibfnamefont {P.~A.}\ \bibnamefont {Davidson}},\ }\href@noop {} {\emph {\bibinfo {title} {Turbulence in rotating, stratified and electrically conducting fluids}}}\ (\bibinfo  {publisher} {Cambridge University Press},\ \bibinfo {year} {2013})\BibitemShut {NoStop}%
\bibitem [{\citenamefont {Davidson}(2015)}]{davidson2015turbulence}%
  \BibitemOpen
  \bibfield  {author} {\bibinfo {author} {\bibfnamefont {P.~A.}\ \bibnamefont {Davidson}},\ }\href@noop {} {\emph {\bibinfo {title} {Turbulence: an introduction for scientists and engineers}}}\ (\bibinfo  {publisher} {Oxford university press},\ \bibinfo {year} {2015})\BibitemShut {NoStop}%
\bibitem [{\citenamefont {Alexakis}\ and\ \citenamefont {Biferale}(2018)}]{alexakis2018cascades}%
  \BibitemOpen
  \bibfield  {author} {\bibinfo {author} {\bibfnamefont {A.}~\bibnamefont {Alexakis}}\ and\ \bibinfo {author} {\bibfnamefont {L.}~\bibnamefont {Biferale}},\ }\bibfield  {title} {\bibinfo {title} {Cascades and transitions in turbulent flows},\ }\href@noop {} {\bibfield  {journal} {\bibinfo  {journal} {Physics Reports}\ }\textbf {\bibinfo {volume} {767}},\ \bibinfo {pages} {1} (\bibinfo {year} {2018})}\BibitemShut {NoStop}%
\bibitem [{\citenamefont {Scott}(2014)}]{scott2014wave}%
  \BibitemOpen
  \bibfield  {author} {\bibinfo {author} {\bibfnamefont {J.~F.}\ \bibnamefont {Scott}},\ }\bibfield  {title} {\bibinfo {title} {Wave turbulence in a rotating channel},\ }\href@noop {} {\bibfield  {journal} {\bibinfo  {journal} {Journal of fluid mechanics}\ }\textbf {\bibinfo {volume} {741}},\ \bibinfo {pages} {316} (\bibinfo {year} {2014})}\BibitemShut {NoStop}%
\bibitem [{\citenamefont {van Kan}\ and\ \citenamefont {Alexakis}(2020)}]{van2020critical}%
  \BibitemOpen
  \bibfield  {author} {\bibinfo {author} {\bibfnamefont {A.}~\bibnamefont {van Kan}}\ and\ \bibinfo {author} {\bibfnamefont {A.}~\bibnamefont {Alexakis}},\ }\bibfield  {title} {\bibinfo {title} {Critical transition in fast-rotating turbulence within highly elongated domains},\ }\href@noop {} {\bibfield  {journal} {\bibinfo  {journal} {Journal of Fluid Mechanics}\ }\textbf {\bibinfo {volume} {899}} (\bibinfo {year} {2020})}\BibitemShut {NoStop}%
\bibitem [{\citenamefont {Greenspan}\ \emph {et~al.}(1968)\citenamefont {Greenspan} \emph {et~al.}}]{greenspan1968theory}%
  \BibitemOpen
  \bibfield  {author} {\bibinfo {author} {\bibfnamefont {H.~P.}\ \bibnamefont {Greenspan}} \emph {et~al.},\ }\href@noop {} {\emph {\bibinfo {title} {The theory of rotating fluids}}}\ (\bibinfo  {publisher} {CUP Archive},\ \bibinfo {year} {1968})\BibitemShut {NoStop}%
\bibitem [{dis()}]{disspersion_symmetry}%
  \BibitemOpen
  \href@noop {} {}\bibinfo {note} {The dispersion relation for inertial waves has a symmetry with respect to the transformation $\omega \rightarrow -\omega$ and $k_z \rightarrow -k_z$, so it is redundant to calculate the distribution of negative frequencies in time. we will assume that $k_z$ can be negative and $\omega$ is positive.}\BibitemShut {Stop}%
\bibitem [{\citenamefont {Bordes}\ \emph {et~al.}(2012)\citenamefont {Bordes}, \citenamefont {Moisy}, \citenamefont {Dauxois},\ and\ \citenamefont {Cortet}}]{bordes2012experimental}%
  \BibitemOpen
  \bibfield  {author} {\bibinfo {author} {\bibfnamefont {G.}~\bibnamefont {Bordes}}, \bibinfo {author} {\bibfnamefont {F.}~\bibnamefont {Moisy}}, \bibinfo {author} {\bibfnamefont {T.}~\bibnamefont {Dauxois}},\ and\ \bibinfo {author} {\bibfnamefont {P.-P.}\ \bibnamefont {Cortet}},\ }\bibfield  {title} {\bibinfo {title} {Experimental evidence of a triadic resonance of plane inertial waves in a rotating fluid},\ }\href@noop {} {\bibfield  {journal} {\bibinfo  {journal} {Physics of Fluids}\ }\textbf {\bibinfo {volume} {24}},\ \bibinfo {pages} {014105} (\bibinfo {year} {2012})}\BibitemShut {NoStop}%
\bibitem [{\citenamefont {Brunet}\ \emph {et~al.}(2020)\citenamefont {Brunet}, \citenamefont {Gallet},\ and\ \citenamefont {Cortet}}]{cortet2020shortcut}%
  \BibitemOpen
  \bibfield  {author} {\bibinfo {author} {\bibfnamefont {M.}~\bibnamefont {Brunet}}, \bibinfo {author} {\bibfnamefont {B.}~\bibnamefont {Gallet}},\ and\ \bibinfo {author} {\bibfnamefont {P.-P.}\ \bibnamefont {Cortet}},\ }\bibfield  {title} {\bibinfo {title} {Shortcut to geostrophy in wave-driven rotating turbulence: the quartetic instability},\ }\href@noop {} {\bibfield  {journal} {\bibinfo  {journal} {Physical Review Letters}\ }\textbf {\bibinfo {volume} {124}},\ \bibinfo {pages} {124501} (\bibinfo {year} {2020})}\BibitemShut {NoStop}%
\bibitem [{\citenamefont {Yarom}\ and\ \citenamefont {Sharon}(2014)}]{yarom2014experimental}%
  \BibitemOpen
  \bibfield  {author} {\bibinfo {author} {\bibfnamefont {E.}~\bibnamefont {Yarom}}\ and\ \bibinfo {author} {\bibfnamefont {E.}~\bibnamefont {Sharon}},\ }\bibfield  {title} {\bibinfo {title} {Experimental observation of steady inertial wave turbulence in deep rotating flows},\ }\href@noop {} {\bibfield  {journal} {\bibinfo  {journal} {Nature Physics}\ }\textbf {\bibinfo {volume} {10}},\ \bibinfo {pages} {510} (\bibinfo {year} {2014})}\BibitemShut {NoStop}%
\bibitem [{\citenamefont {Campagne}\ \emph {et~al.}(2015)\citenamefont {Campagne}, \citenamefont {Gallet}, \citenamefont {Moisy},\ and\ \citenamefont {Cortet}}]{cortetcampagne2015disentangling}%
  \BibitemOpen
  \bibfield  {author} {\bibinfo {author} {\bibfnamefont {A.}~\bibnamefont {Campagne}}, \bibinfo {author} {\bibfnamefont {B.}~\bibnamefont {Gallet}}, \bibinfo {author} {\bibfnamefont {F.}~\bibnamefont {Moisy}},\ and\ \bibinfo {author} {\bibfnamefont {P.-P.}\ \bibnamefont {Cortet}},\ }\bibfield  {title} {\bibinfo {title} {Disentangling inertial waves from eddy turbulence in a forced rotating-turbulence experiment},\ }\href@noop {} {\bibfield  {journal} {\bibinfo  {journal} {Physical Review E}\ }\textbf {\bibinfo {volume} {91}},\ \bibinfo {pages} {043016} (\bibinfo {year} {2015})}\BibitemShut {NoStop}%
\bibitem [{\citenamefont {Duran-Matute}\ \emph {et~al.}(2013)\citenamefont {Duran-Matute}, \citenamefont {Fl{\'o}r}, \citenamefont {Godeferd},\ and\ \citenamefont {Jause-Labert}}]{duran2013turbulence}%
  \BibitemOpen
  \bibfield  {author} {\bibinfo {author} {\bibfnamefont {M.}~\bibnamefont {Duran-Matute}}, \bibinfo {author} {\bibfnamefont {J.-B.}\ \bibnamefont {Fl{\'o}r}}, \bibinfo {author} {\bibfnamefont {F.~S.}\ \bibnamefont {Godeferd}},\ and\ \bibinfo {author} {\bibfnamefont {C.}~\bibnamefont {Jause-Labert}},\ }\bibfield  {title} {\bibinfo {title} {Turbulence and columnar vortex formation through inertial-wave focusing},\ }\href@noop {} {\bibfield  {journal} {\bibinfo  {journal} {Physical Review E}\ }\textbf {\bibinfo {volume} {87}},\ \bibinfo {pages} {041001} (\bibinfo {year} {2013})}\BibitemShut {NoStop}%
\bibitem [{\citenamefont {Cambon}\ \emph {et~al.}(1997)\citenamefont {Cambon}, \citenamefont {Mansour},\ and\ \citenamefont {Godeferd}}]{cambon1997energy}%
  \BibitemOpen
  \bibfield  {author} {\bibinfo {author} {\bibfnamefont {C.}~\bibnamefont {Cambon}}, \bibinfo {author} {\bibfnamefont {N.~N.}\ \bibnamefont {Mansour}},\ and\ \bibinfo {author} {\bibfnamefont {F.~S.}\ \bibnamefont {Godeferd}},\ }\bibfield  {title} {\bibinfo {title} {Energy transfer in rotating turbulence},\ }\href@noop {} {\bibfield  {journal} {\bibinfo  {journal} {Journal of Fluid Mechanics}\ }\textbf {\bibinfo {volume} {337}},\ \bibinfo {pages} {303} (\bibinfo {year} {1997})}\BibitemShut {NoStop}%
\bibitem [{\citenamefont {Godeferd}\ and\ \citenamefont {Moisy}(2015)}]{godeferd2015structure}%
  \BibitemOpen
  \bibfield  {author} {\bibinfo {author} {\bibfnamefont {F.~S.}\ \bibnamefont {Godeferd}}\ and\ \bibinfo {author} {\bibfnamefont {F.}~\bibnamefont {Moisy}},\ }\bibfield  {title} {\bibinfo {title} {Structure and dynamics of rotating turbulence: a review of recent experimental and numerical results},\ }\href@noop {} {\bibfield  {journal} {\bibinfo  {journal} {Applied Mechanics Reviews}\ }\textbf {\bibinfo {volume} {67}} (\bibinfo {year} {2015})}\BibitemShut {NoStop}%
\bibitem [{\citenamefont {Kolvin}\ \emph {et~al.}(2009)\citenamefont {Kolvin}, \citenamefont {Cohen}, \citenamefont {Vardi},\ and\ \citenamefont {Sharon}}]{kolvin2009energy}%
  \BibitemOpen
  \bibfield  {author} {\bibinfo {author} {\bibfnamefont {I.}~\bibnamefont {Kolvin}}, \bibinfo {author} {\bibfnamefont {K.}~\bibnamefont {Cohen}}, \bibinfo {author} {\bibfnamefont {Y.}~\bibnamefont {Vardi}},\ and\ \bibinfo {author} {\bibfnamefont {E.}~\bibnamefont {Sharon}},\ }\bibfield  {title} {\bibinfo {title} {Energy transfer by inertial waves during the buildup of turbulence in a rotating system},\ }\href@noop {} {\bibfield  {journal} {\bibinfo  {journal} {Physical Review Letters}\ }\textbf {\bibinfo {volume} {102}},\ \bibinfo {pages} {014503} (\bibinfo {year} {2009})}\BibitemShut {NoStop}%
\bibitem [{\citenamefont {Davidson}\ \emph {et~al.}(2006)\citenamefont {Davidson}, \citenamefont {Staplehurst},\ and\ \citenamefont {Dalziel}}]{davidson2006evolution}%
  \BibitemOpen
  \bibfield  {author} {\bibinfo {author} {\bibfnamefont {P.}~\bibnamefont {Davidson}}, \bibinfo {author} {\bibfnamefont {P.}~\bibnamefont {Staplehurst}},\ and\ \bibinfo {author} {\bibfnamefont {S.}~\bibnamefont {Dalziel}},\ }\bibfield  {title} {\bibinfo {title} {On the evolution of eddies in a rapidly rotating system},\ }\href@noop {} {\bibfield  {journal} {\bibinfo  {journal} {Journal of Fluid Mechanics}\ }\textbf {\bibinfo {volume} {557}},\ \bibinfo {pages} {135} (\bibinfo {year} {2006})}\BibitemShut {NoStop}%
\bibitem [{\citenamefont {Staplehurst}\ \emph {et~al.}(2008)\citenamefont {Staplehurst}, \citenamefont {Davidson},\ and\ \citenamefont {Dalziel}}]{staplehurst2008structure}%
  \BibitemOpen
  \bibfield  {author} {\bibinfo {author} {\bibfnamefont {P.}~\bibnamefont {Staplehurst}}, \bibinfo {author} {\bibfnamefont {P.}~\bibnamefont {Davidson}},\ and\ \bibinfo {author} {\bibfnamefont {S.}~\bibnamefont {Dalziel}},\ }\bibfield  {title} {\bibinfo {title} {Structure formation in homogeneous freely decaying rotating turbulence},\ }\href@noop {} {\bibfield  {journal} {\bibinfo  {journal} {Journal of Fluid Mechanics}\ }\textbf {\bibinfo {volume} {598}},\ \bibinfo {pages} {81} (\bibinfo {year} {2008})}\BibitemShut {NoStop}%
\bibitem [{\citenamefont {Bewley}\ \emph {et~al.}(2007)\citenamefont {Bewley}, \citenamefont {Lathrop}, \citenamefont {Maas},\ and\ \citenamefont {Sreenivasan}}]{bewley2007inertial}%
  \BibitemOpen
  \bibfield  {author} {\bibinfo {author} {\bibfnamefont {G.~P.}\ \bibnamefont {Bewley}}, \bibinfo {author} {\bibfnamefont {D.~P.}\ \bibnamefont {Lathrop}}, \bibinfo {author} {\bibfnamefont {L.~R.}\ \bibnamefont {Maas}},\ and\ \bibinfo {author} {\bibfnamefont {K.}~\bibnamefont {Sreenivasan}},\ }\bibfield  {title} {\bibinfo {title} {Inertial waves in rotating grid turbulence},\ }\href@noop {} {\bibfield  {journal} {\bibinfo  {journal} {Physics of Fluids}\ }\textbf {\bibinfo {volume} {19}},\ \bibinfo {pages} {071701} (\bibinfo {year} {2007})}\BibitemShut {NoStop}%
\bibitem [{\citenamefont {Morize}\ \emph {et~al.}(2005)\citenamefont {Morize}, \citenamefont {Moisy},\ and\ \citenamefont {Rabaud}}]{morize2005decaying}%
  \BibitemOpen
  \bibfield  {author} {\bibinfo {author} {\bibfnamefont {C.}~\bibnamefont {Morize}}, \bibinfo {author} {\bibfnamefont {F.}~\bibnamefont {Moisy}},\ and\ \bibinfo {author} {\bibfnamefont {M.}~\bibnamefont {Rabaud}},\ }\bibfield  {title} {\bibinfo {title} {Decaying grid-generated turbulence in a rotating tank},\ }\href@noop {} {\bibfield  {journal} {\bibinfo  {journal} {Physics of fluids}\ }\textbf {\bibinfo {volume} {17}},\ \bibinfo {pages} {095105} (\bibinfo {year} {2005})}\BibitemShut {NoStop}%
\bibitem [{\citenamefont {Galtier}(2003)}]{galtier2003weak}%
  \BibitemOpen
  \bibfield  {author} {\bibinfo {author} {\bibfnamefont {S.}~\bibnamefont {Galtier}},\ }\bibfield  {title} {\bibinfo {title} {Weak inertial-wave turbulence theory},\ }\href@noop {} {\bibfield  {journal} {\bibinfo  {journal} {Physical Review E}\ }\textbf {\bibinfo {volume} {68}},\ \bibinfo {pages} {015301} (\bibinfo {year} {2003})}\BibitemShut {NoStop}%
\bibitem [{\citenamefont {Nazarenko}(2011)}]{nazarenko2011wave}%
  \BibitemOpen
  \bibfield  {author} {\bibinfo {author} {\bibfnamefont {S.}~\bibnamefont {Nazarenko}},\ }\href@noop {} {\emph {\bibinfo {title} {Wave turbulence}}},\ Vol.\ \bibinfo {volume} {825}\ (\bibinfo  {publisher} {Springer Science \& Business Media},\ \bibinfo {year} {2011})\BibitemShut {NoStop}%
\bibitem [{\citenamefont {Smith}\ and\ \citenamefont {Waleffe}(1999)}]{smith1999transfer}%
  \BibitemOpen
  \bibfield  {author} {\bibinfo {author} {\bibfnamefont {L.~M.}\ \bibnamefont {Smith}}\ and\ \bibinfo {author} {\bibfnamefont {F.}~\bibnamefont {Waleffe}},\ }\bibfield  {title} {\bibinfo {title} {Transfer of energy to two-dimensional large scales in forced, rotating three-dimensional turbulence},\ }\href@noop {} {\bibfield  {journal} {\bibinfo  {journal} {Physics of fluids}\ }\textbf {\bibinfo {volume} {11}},\ \bibinfo {pages} {1608} (\bibinfo {year} {1999})}\BibitemShut {NoStop}%
\bibitem [{\citenamefont {Monsalve}\ \emph {et~al.}(2020)\citenamefont {Monsalve}, \citenamefont {Brunet}, \citenamefont {Gallet},\ and\ \citenamefont {Cortet}}]{cortet2020quantitative}%
  \BibitemOpen
  \bibfield  {author} {\bibinfo {author} {\bibfnamefont {E.}~\bibnamefont {Monsalve}}, \bibinfo {author} {\bibfnamefont {M.}~\bibnamefont {Brunet}}, \bibinfo {author} {\bibfnamefont {B.}~\bibnamefont {Gallet}},\ and\ \bibinfo {author} {\bibfnamefont {P.-P.}\ \bibnamefont {Cortet}},\ }\bibfield  {title} {\bibinfo {title} {Quantitative experimental observation of weak inertial-wave turbulence},\ }\href@noop {} {\bibfield  {journal} {\bibinfo  {journal} {Physical Review Letters}\ }\textbf {\bibinfo {volume} {125}},\ \bibinfo {pages} {254502} (\bibinfo {year} {2020})}\BibitemShut {NoStop}%
\bibitem [{sup()}]{supp_material}%
  \BibitemOpen
  \href@noop {} {\bibinfo {title} {Supplemental information}},\ \bibinfo {howpublished} {See Supplemental Information at URL will be inserted by publisher.}\BibitemShut {Stop}%
\bibitem [{\citenamefont {Yarom}\ \emph {et~al.}(2013)\citenamefont {Yarom}, \citenamefont {Vardi},\ and\ \citenamefont {Sharon}}]{yarom2013experimental}%
  \BibitemOpen
  \bibfield  {author} {\bibinfo {author} {\bibfnamefont {E.}~\bibnamefont {Yarom}}, \bibinfo {author} {\bibfnamefont {Y.}~\bibnamefont {Vardi}},\ and\ \bibinfo {author} {\bibfnamefont {E.}~\bibnamefont {Sharon}},\ }\bibfield  {title} {\bibinfo {title} {Experimental quantification of inverse energy cascade in deep rotating turbulence},\ }\href@noop {} {\bibfield  {journal} {\bibinfo  {journal} {Physics of Fluids}\ }\textbf {\bibinfo {volume} {25}},\ \bibinfo {pages} {085105} (\bibinfo {year} {2013})}\BibitemShut {NoStop}%
\bibitem [{\citenamefont {Buzzicotti}\ \emph {et~al.}(2018)\citenamefont {Buzzicotti}, \citenamefont {Clark Di~Leoni},\ and\ \citenamefont {Biferale}}]{buzzicotti2018inverse}%
  \BibitemOpen
  \bibfield  {author} {\bibinfo {author} {\bibfnamefont {M.}~\bibnamefont {Buzzicotti}}, \bibinfo {author} {\bibfnamefont {P.}~\bibnamefont {Clark Di~Leoni}},\ and\ \bibinfo {author} {\bibfnamefont {L.}~\bibnamefont {Biferale}},\ }\bibfield  {title} {\bibinfo {title} {On the inverse energy transfer in rotating turbulence},\ }\href@noop {} {\bibfield  {journal} {\bibinfo  {journal} {The European Physical Journal E}\ }\textbf {\bibinfo {volume} {41}},\ \bibinfo {pages} {1} (\bibinfo {year} {2018})}\BibitemShut {NoStop}%
\bibitem [{\citenamefont {Sen}\ \emph {et~al.}(2012)\citenamefont {Sen}, \citenamefont {Mininni}, \citenamefont {Rosenberg},\ and\ \citenamefont {Pouquet}}]{sen2012anisotropy}%
  \BibitemOpen
  \bibfield  {author} {\bibinfo {author} {\bibfnamefont {A.}~\bibnamefont {Sen}}, \bibinfo {author} {\bibfnamefont {P.~D.}\ \bibnamefont {Mininni}}, \bibinfo {author} {\bibfnamefont {D.}~\bibnamefont {Rosenberg}},\ and\ \bibinfo {author} {\bibfnamefont {A.}~\bibnamefont {Pouquet}},\ }\bibfield  {title} {\bibinfo {title} {Anisotropy and nonuniversality in scaling laws of the large-scale energy spectrum in rotating turbulence},\ }\href@noop {} {\bibfield  {journal} {\bibinfo  {journal} {Physical Review E}\ }\textbf {\bibinfo {volume} {86}},\ \bibinfo {pages} {036319} (\bibinfo {year} {2012})}\BibitemShut {NoStop}%
\bibitem [{\citenamefont {Baroud}\ \emph {et~al.}(2003)\citenamefont {Baroud}, \citenamefont {Plapp}, \citenamefont {Swinney},\ and\ \citenamefont {She}}]{baroud2003scaling}%
  \BibitemOpen
  \bibfield  {author} {\bibinfo {author} {\bibfnamefont {C.~N.}\ \bibnamefont {Baroud}}, \bibinfo {author} {\bibfnamefont {B.~B.}\ \bibnamefont {Plapp}}, \bibinfo {author} {\bibfnamefont {H.~L.}\ \bibnamefont {Swinney}},\ and\ \bibinfo {author} {\bibfnamefont {Z.-S.}\ \bibnamefont {She}},\ }\bibfield  {title} {\bibinfo {title} {Scaling in three-dimensional and quasi-two-dimensional rotating turbulent flows},\ }\href@noop {} {\bibfield  {journal} {\bibinfo  {journal} {Physics of Fluids}\ }\textbf {\bibinfo {volume} {15}},\ \bibinfo {pages} {2091} (\bibinfo {year} {2003})}\BibitemShut {NoStop}%
\bibitem [{\citenamefont {Lamriben}\ \emph {et~al.}(2011)\citenamefont {Lamriben}, \citenamefont {Cortet},\ and\ \citenamefont {Moisy}}]{lamriben2011direct}%
  \BibitemOpen
  \bibfield  {author} {\bibinfo {author} {\bibfnamefont {C.}~\bibnamefont {Lamriben}}, \bibinfo {author} {\bibfnamefont {P.-P.}\ \bibnamefont {Cortet}},\ and\ \bibinfo {author} {\bibfnamefont {F.}~\bibnamefont {Moisy}},\ }\bibfield  {title} {\bibinfo {title} {Direct measurements of anisotropic energy transfers in a rotating turbulence experiment},\ }\href@noop {} {\bibfield  {journal} {\bibinfo  {journal} {Physical review letters}\ }\textbf {\bibinfo {volume} {107}},\ \bibinfo {pages} {024503} (\bibinfo {year} {2011})}\BibitemShut {NoStop}%
\bibitem [{\citenamefont {Campagne}\ \emph {et~al.}(2014)\citenamefont {Campagne}, \citenamefont {Gallet}, \citenamefont {Moisy},\ and\ \citenamefont {Cortet}}]{campagne2014direct}%
  \BibitemOpen
  \bibfield  {author} {\bibinfo {author} {\bibfnamefont {A.}~\bibnamefont {Campagne}}, \bibinfo {author} {\bibfnamefont {B.}~\bibnamefont {Gallet}}, \bibinfo {author} {\bibfnamefont {F.}~\bibnamefont {Moisy}},\ and\ \bibinfo {author} {\bibfnamefont {P.-P.}\ \bibnamefont {Cortet}},\ }\bibfield  {title} {\bibinfo {title} {Direct and inverse energy cascades in a forced rotating turbulence experiment},\ }\href@noop {} {\bibfield  {journal} {\bibinfo  {journal} {Physics of Fluids}\ }\textbf {\bibinfo {volume} {26}},\ \bibinfo {pages} {125112} (\bibinfo {year} {2014})}\BibitemShut {NoStop}%
\bibitem [{\citenamefont {di~Leoni}\ and\ \citenamefont {Mininni}(2016)}]{di2016quantifying}%
  \BibitemOpen
  \bibfield  {author} {\bibinfo {author} {\bibfnamefont {P.~C.}\ \bibnamefont {di~Leoni}}\ and\ \bibinfo {author} {\bibfnamefont {P.~D.}\ \bibnamefont {Mininni}},\ }\bibfield  {title} {\bibinfo {title} {Quantifying resonant and near-resonant interactions in rotating turbulence},\ }\href@noop {} {\bibfield  {journal} {\bibinfo  {journal} {Journal of Fluid Mechanics}\ }\textbf {\bibinfo {volume} {809}},\ \bibinfo {pages} {821} (\bibinfo {year} {2016})}\BibitemShut {NoStop}%
\bibitem [{\citenamefont {Le~Reun}\ \emph {et~al.}(2020)\citenamefont {Le~Reun}, \citenamefont {Gallet}, \citenamefont {Favier},\ and\ \citenamefont {Le~Bars}}]{le2020near}%
  \BibitemOpen
  \bibfield  {author} {\bibinfo {author} {\bibfnamefont {T.}~\bibnamefont {Le~Reun}}, \bibinfo {author} {\bibfnamefont {B.}~\bibnamefont {Gallet}}, \bibinfo {author} {\bibfnamefont {B.}~\bibnamefont {Favier}},\ and\ \bibinfo {author} {\bibfnamefont {M.}~\bibnamefont {Le~Bars}},\ }\bibfield  {title} {\bibinfo {title} {Near-resonant instability of geostrophic modes: beyond greenspan's theorem},\ }\href@noop {} {\bibfield  {journal} {\bibinfo  {journal} {Journal of Fluid Mechanics}\ }\textbf {\bibinfo {volume} {900}} (\bibinfo {year} {2020})}\BibitemShut {NoStop}%
\bibitem [{\citenamefont {Bellet}\ \emph {et~al.}(2006)\citenamefont {Bellet}, \citenamefont {Godeferd}, \citenamefont {Scott},\ and\ \citenamefont {Cambon}}]{bellet2006wave}%
  \BibitemOpen
  \bibfield  {author} {\bibinfo {author} {\bibfnamefont {F.}~\bibnamefont {Bellet}}, \bibinfo {author} {\bibfnamefont {F.}~\bibnamefont {Godeferd}}, \bibinfo {author} {\bibfnamefont {J.}~\bibnamefont {Scott}},\ and\ \bibinfo {author} {\bibfnamefont {C.}~\bibnamefont {Cambon}},\ }\bibfield  {title} {\bibinfo {title} {Wave turbulence in rapidly rotating flows},\ }\href@noop {} {\bibfield  {journal} {\bibinfo  {journal} {Journal of Fluid Mechanics}\ }\textbf {\bibinfo {volume} {562}},\ \bibinfo {pages} {83} (\bibinfo {year} {2006})}\BibitemShut {NoStop}%
\bibitem [{\citenamefont {Gallet}(2015)}]{gallet2015exact}%
  \BibitemOpen
  \bibfield  {author} {\bibinfo {author} {\bibfnamefont {B.}~\bibnamefont {Gallet}},\ }\bibfield  {title} {\bibinfo {title} {Exact two-dimensionalization of rapidly rotating large-reynolds-number flows},\ }\href@noop {} {\bibfield  {journal} {\bibinfo  {journal} {Journal of Fluid Mechanics}\ }\textbf {\bibinfo {volume} {783}},\ \bibinfo {pages} {412} (\bibinfo {year} {2015})}\BibitemShut {NoStop}%
\bibitem [{\citenamefont {Nazarenko}\ and\ \citenamefont {Schekochihin}(2011)}]{nazarenko2011critical}%
  \BibitemOpen
  \bibfield  {author} {\bibinfo {author} {\bibfnamefont {S.~V.}\ \bibnamefont {Nazarenko}}\ and\ \bibinfo {author} {\bibfnamefont {A.~A.}\ \bibnamefont {Schekochihin}},\ }\bibfield  {title} {\bibinfo {title} {Critical balance in magnetohydrodynamic, rotating and stratified turbulence: towards a universal scaling conjecture},\ }\href@noop {} {\bibfield  {journal} {\bibinfo  {journal} {Journal of Fluid Mechanics}\ }\textbf {\bibinfo {volume} {677}},\ \bibinfo {pages} {134} (\bibinfo {year} {2011})}\BibitemShut {NoStop}%
\bibitem [{\citenamefont {Yarom}\ \emph {et~al.}(2017)\citenamefont {Yarom}, \citenamefont {Salhov},\ and\ \citenamefont {Sharon}}]{yarom2017experimental}%
  \BibitemOpen
  \bibfield  {author} {\bibinfo {author} {\bibfnamefont {E.}~\bibnamefont {Yarom}}, \bibinfo {author} {\bibfnamefont {A.}~\bibnamefont {Salhov}},\ and\ \bibinfo {author} {\bibfnamefont {E.}~\bibnamefont {Sharon}},\ }\bibfield  {title} {\bibinfo {title} {Experimental quantification of nonlinear time scales in inertial wave rotating turbulence},\ }\href@noop {} {\bibfield  {journal} {\bibinfo  {journal} {Physical Review Fluids}\ }\textbf {\bibinfo {volume} {2}},\ \bibinfo {pages} {122601} (\bibinfo {year} {2017})}\BibitemShut {NoStop}%
\bibitem [{\citenamefont {Le~Reun}\ \emph {et~al.}(2017)\citenamefont {Le~Reun}, \citenamefont {Favier}, \citenamefont {Barker},\ and\ \citenamefont {Le~Bars}}]{le2017inertial}%
  \BibitemOpen
  \bibfield  {author} {\bibinfo {author} {\bibfnamefont {T.}~\bibnamefont {Le~Reun}}, \bibinfo {author} {\bibfnamefont {B.}~\bibnamefont {Favier}}, \bibinfo {author} {\bibfnamefont {A.~J.}\ \bibnamefont {Barker}},\ and\ \bibinfo {author} {\bibfnamefont {M.}~\bibnamefont {Le~Bars}},\ }\bibfield  {title} {\bibinfo {title} {Inertial wave turbulence driven by elliptical instability},\ }\href@noop {} {\bibfield  {journal} {\bibinfo  {journal} {Physical Review Letters}\ }\textbf {\bibinfo {volume} {119}},\ \bibinfo {pages} {034502} (\bibinfo {year} {2017})}\BibitemShut {NoStop}%
\bibitem [{\citenamefont {Clark Di~Leoni}\ \emph {et~al.}(2014)\citenamefont {Clark Di~Leoni}, \citenamefont {Cobelli}, \citenamefont {Mininni}, \citenamefont {Dmitruk},\ and\ \citenamefont {Matthaeus}}]{clark2014quantification}%
  \BibitemOpen
  \bibfield  {author} {\bibinfo {author} {\bibfnamefont {P.}~\bibnamefont {Clark Di~Leoni}}, \bibinfo {author} {\bibfnamefont {P.~J.}\ \bibnamefont {Cobelli}}, \bibinfo {author} {\bibfnamefont {P.~D.}\ \bibnamefont {Mininni}}, \bibinfo {author} {\bibfnamefont {P.}~\bibnamefont {Dmitruk}},\ and\ \bibinfo {author} {\bibfnamefont {W.}~\bibnamefont {Matthaeus}},\ }\bibfield  {title} {\bibinfo {title} {Quantification of the strength of inertial waves in a rotating turbulent flow},\ }\href@noop {} {\bibfield  {journal} {\bibinfo  {journal} {Physics of Fluids}\ }\textbf {\bibinfo {volume} {26}},\ \bibinfo {pages} {035106} (\bibinfo {year} {2014})}\BibitemShut {NoStop}%
\bibitem [{\citenamefont {Salhov}\ \emph {et~al.}(2019)\citenamefont {Salhov}, \citenamefont {Yarom},\ and\ \citenamefont {Sharon}}]{salhov2019measurements}%
  \BibitemOpen
  \bibfield  {author} {\bibinfo {author} {\bibfnamefont {A.}~\bibnamefont {Salhov}}, \bibinfo {author} {\bibfnamefont {E.}~\bibnamefont {Yarom}},\ and\ \bibinfo {author} {\bibfnamefont {E.}~\bibnamefont {Sharon}},\ }\bibfield  {title} {\bibinfo {title} {Measurements of inertial wave packets propagating within steady rotating turbulence},\ }\href@noop {} {\bibfield  {journal} {\bibinfo  {journal} {EPL (Europhysics Letters)}\ }\textbf {\bibinfo {volume} {125}},\ \bibinfo {pages} {24003} (\bibinfo {year} {2019})}\BibitemShut {NoStop}%
\bibitem [{vid(2023{\natexlab{a}})}]{video_ref2}%
  \BibitemOpen
  \href {URL will be inserted by publisher} {}\bibinfo {howpublished} {Online video: Energy density in time, shown on three planes of the volume and on a vertical plane} (\bibinfo {year} {2023}{\natexlab{a}})\BibitemShut {NoStop}%
\bibitem [{vid(2023{\natexlab{b}})}]{video_ref}%
  \BibitemOpen
  \href {URL will be inserted by publisher.} {}\bibinfo {howpublished} {Online video: excess energy spectrum, on the $\omega-\theta$ plane} (\bibinfo {year} {2023}{\natexlab{b}})\BibitemShut {NoStop}%
\bibitem [{\citenamefont {Kraichnan}(1967)}]{kraichnan1967inertial}%
  \BibitemOpen
  \bibfield  {author} {\bibinfo {author} {\bibfnamefont {R.~H.}\ \bibnamefont {Kraichnan}},\ }\bibfield  {title} {\bibinfo {title} {Inertial ranges in two-dimensional turbulence},\ }\href@noop {} {\bibfield  {journal} {\bibinfo  {journal} {The Physics of Fluids}\ }\textbf {\bibinfo {volume} {10}},\ \bibinfo {pages} {1417} (\bibinfo {year} {1967})}\BibitemShut {NoStop}%
\bibitem [{\citenamefont {Frisch}\ and\ \citenamefont {Sulem}(1984)}]{frisch1984numerical}%
  \BibitemOpen
  \bibfield  {author} {\bibinfo {author} {\bibfnamefont {U.}~\bibnamefont {Frisch}}\ and\ \bibinfo {author} {\bibfnamefont {P.-L.}\ \bibnamefont {Sulem}},\ }\bibfield  {title} {\bibinfo {title} {Numerical simulation of the inverse cascade in two-dimensional turbulence},\ }\href@noop {} {\bibfield  {journal} {\bibinfo  {journal} {The Physics of fluids}\ }\textbf {\bibinfo {volume} {27}},\ \bibinfo {pages} {1921} (\bibinfo {year} {1984})}\BibitemShut {NoStop}%
\bibitem [{\citenamefont {Tabeling}(2002)}]{tabeling2002two}%
  \BibitemOpen
  \bibfield  {author} {\bibinfo {author} {\bibfnamefont {P.}~\bibnamefont {Tabeling}},\ }\bibfield  {title} {\bibinfo {title} {Two-dimensional turbulence: a physicist approach},\ }\href@noop {} {\bibfield  {journal} {\bibinfo  {journal} {Physics reports}\ }\textbf {\bibinfo {volume} {362}},\ \bibinfo {pages} {1} (\bibinfo {year} {2002})}\BibitemShut {NoStop}%
\bibitem [{\citenamefont {Paret}\ and\ \citenamefont {Tabeling}(1997)}]{paret1997experimental}%
  \BibitemOpen
  \bibfield  {author} {\bibinfo {author} {\bibfnamefont {J.}~\bibnamefont {Paret}}\ and\ \bibinfo {author} {\bibfnamefont {P.}~\bibnamefont {Tabeling}},\ }\bibfield  {title} {\bibinfo {title} {Experimental observation of the two-dimensional inverse energy cascade},\ }\href@noop {} {\bibfield  {journal} {\bibinfo  {journal} {Physical review letters}\ }\textbf {\bibinfo {volume} {79}},\ \bibinfo {pages} {4162} (\bibinfo {year} {1997})}\BibitemShut {NoStop}%
\end{thebibliography}%

\end{document}